\documentclass[aps,prl,showpacs,twocolumn,superscriptaddress]{revtex4-1}

\usepackage{amsmath}
\usepackage{amsfonts}
\usepackage{amssymb}
\usepackage{epsfig}

\usepackage{braket}

\usepackage{graphicx}

\usepackage{hyphenat}
\usepackage{hyperref}
\usepackage{placeins}	
\usepackage{color}

\usepackage{array}
\newcolumntype{C}[1]{>{\centering\arraybackslash}m{#1}}
\usepackage{overpic}

\renewcommand{\vec}{\mathbf}

\usepackage[normalem]{ulem}

\usepackage{booktabs}
\usepackage{diagbox}
\usepackage{siunitx}
\AtBeginDocument{
	\heavyrulewidth=.08em
	\lightrulewidth=.05em
	\cmidrulewidth=.03em
	\belowrulesep=.65ex
	\belowbottomsep=0pt
	\aboverulesep=.4ex
	\abovetopsep=0pt
	\cmidrulesep=\doublerulesep
	\cmidrulekern=.5em
	\defaultaddspace=.5em
}
\begin{document}

\title{Quantum criticality of two-dimensional quantum magnets with long-range interactions}
\author{Sebastian Fey}
\affiliation{Lehrstuhl f\"ur Theoretische Physik I, Staudtstra{\ss}e 7, Universit\"at Erlangen-N\"urnberg, D-91058 Erlangen, Germany}
\author{Sebastian C. Kapfer}
\affiliation{Lehrstuhl f\"ur Theoretische Physik I, Staudtstra{\ss}e 7, Universit\"at Erlangen-N\"urnberg, D-91058 Erlangen, Germany}
\author{Kai Phillip Schmidt}
\affiliation{Lehrstuhl f\"ur Theoretische Physik I, Staudtstra{\ss}e 7, Universit\"at Erlangen-N\"urnberg, D-91058 Erlangen, Germany}

\begin{abstract}
 We study the critical breakdown of two-dimensional quantum magnets in the presence of algebraically decaying long-range interactions by investigating the transverse-field Ising model on the square and triangular lattice. This is achieved technically by combining perturbative continuous unitary transformations with classical Monte Carlo simulations to extract high-order series for the one-particle excitations in the high-field quantum paramagnet. We find that the unfrustrated systems change from mean-field to nearest-neighbor universality with continuously varying critical exponents, while the system remains in the universality class of the nearest-neighbor model in the frustrated cases independent of the long-range nature of the interaction.
\end{abstract}

\maketitle

%Introduction
%%%%%%%%%%%%%%%%%%%%%%%%%%%%%%%%%%%%%%%%%%%%%%%%%%%%%%%%%%%%%%%%%%%%%%%%%%%%%%%%%%%%%%%%%%%%

The understanding of quantum phase transitions at zero temperature has been an active research field over many decades, since the diverging quantum fluctuations of quantum many-body systems at a quantum critical point lead to intriguing universal behavior giving rise to many fascinating quantum materials with novel collective effects. The physical properties close to a zero-temperature quantum critical point can be classified for most systems by universality classes, which only depend on the dimension and the symmetry of the underlying system. As a consequence, the critical behavior of many physical systems can be described by paradigmatic models for each universality class, which in many cases correspond to interacting spin systems.

One of the most important microscopic models is the ferromagnetic transverse-field Ising model (TFIM). This unfrustrated system realizes a quantum phase transition between a quantum paramagnet and a \(\mathcal{Z}_2\)-symmetry-broken phase for any lattice in any dimension $d$. The corresponding universality class is the one of the classical Ising model in dimension $d+1$. In general, the quantum-critical properties of unfrustrated models with short-range interactions are well understood. The situation becomes more interesting in the presence of frustration where different types of quantum-critical behavior as well as exotic states of quantum matter are known to occur. Important examples in the framework of fully-frustrated TFIMs are the antiferromagnetic TFIM on the triangular and pyrochlore lattice. For the triangular TFIM an order by disorder mechanism gives rise to a ground state where translational symmetry is broken and the universality class of the quantum phase transition is 3D-XY \cite{Blankschtein1984,Moessner2001,Moessner2003,Powalski2013}. In contrast, on the pyrochlore lattice, disorder by disorder leads to a quantum-disordered Coulomb phase in the antiferromagnetic TFIM \cite{Balents2010,Hermele2004,Shannon2012} displaying emergent quantum electrodynamics and the quantum phase transition to the high-field quantum paramagnet is first order \cite{Roechner2016}.   

All of the above systems are restricted to short-range interactions. However, there are many important physical systems where long-range interactions are relevant \cite{Bitko1996,Chakraborty2004,Bramwell2001,Castelnovo2008,Mengotti2009,Lahaye2009,Peter2012,Britton2012,Islam2013,Jurcevic2014,Richerme2014,Mahmoudian2015,Bohnet2016}. Important examples are dipolar interactions between spins in spin-ice materials giving rise to emergent magnetic monopoles \cite{Castelnovo2008}, effective long-range magnetic interactions between zig-zag edges in graphene \cite{Koop2017}, as well as trapped cold-ion systems in quantum optics for which the nature of interactions can be varied flexibly and which have realized the long-range TFIM (lrTFIM) on the triangular lattice \cite{Britton2012, Islam2013, Bohnet2016}.  

The critical behavior of quantum systems with long-range interactions is much less understood. Several studies have focused on the TFIM chain with long-range interactions \cite{Koffel2012,Knap2013,Fey2016,Sun2017,Horita2017}. For a ferromagnetic Ising exchange there are three different regimes. Besides 2D-Ising criticality as for the nearest-neighbor TFIM chain and mean-field (MF) behavior, for intermediate long-range interactions, there is a window with continuously varying critical exponents. In contrast, a recent investigation of the frustrated antiferromagnetic TFIM chain with long-range interactions indicates that the critical behavior is always 2D-Ising independent of the nature of the long-range interaction \cite{Sun2017}. Much less is known in $(2+1)$-dimensions \cite{Humeniuk2016}, since numerical investigations are much harder to perform.  This is especially true when it comes to the interplay of long-range interactions and frustration. In this letter, we combine high-order series expansions with classical Monte Carlo simulations to investigate such interesting and challenging quantum systems.

\begin{figure}[t]
	\centering
	\includegraphics[width=.85\columnwidth]{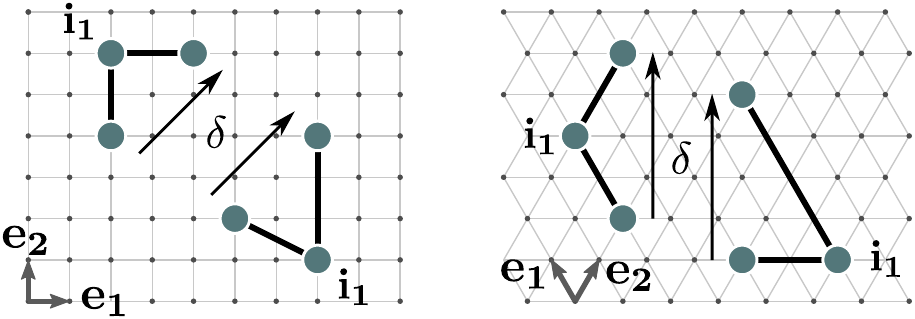}
	\caption{Illustration of the square (left) and triangular (right) lattice with basis vectors $\vec{e_1}$ and $\vec{e_2}$. Various embedding examples of the three-site chain graph for a fixed $\vec\delta$ are shown.}
	\label{fig:graph_embeddings}
\end{figure}

%Model
%%%%%%%%%%%%%%%%%%%%%%%%%%%%%%%%%%%%%%%%%%%%%%%%%%%%%%%%%%%%%%%%%%%%%%%%%%%%%%%%%%%%%%%%%%%%
{\it{Model:}}
We study the lrTFIM given by
\begin{align}
  \mathcal{H} =-\frac{1}{2}\sum_{{\bf j}}\sigma_{{\bf j}}^z -\lambda\sum_{{\bf i}    \neq {\bf j}} \frac{1}{|{\bf i}-{\bf j}|^\alpha}\sigma_{{\bf i}}^x\;\sigma_{{\bf j}}^x~ , \label{eq:H_tfim_orig}
\end{align}
with Pauli matrices \(\sigma_{{\bf i}}^{x/z}\) describing spins-1/2 located on lattice sites \({\bf i}\). Positive (negative) $\lambda$ correspond to (anti)-ferromagnetic interactions. Tuning the positive parameter \(\alpha\) changes the long-range behavior of the interaction, where \(\alpha=\infty\) recovers the nearest-neighbor TFIM. In this work we focus on the square and triangular lattice illustrated in Fig.~\ref{fig:graph_embeddings}.

%Approach
%%%%%%%%%%%%%%%%%%%%%%%%%%%%%%%%%%%%%%%%%%%%%%%%%%%%%%%%%%%%%%%%%%%%%%%%%%%%%%%%%%%%%%%%%%%%
{\it{Approach:}} 
We perform high-order series expansions in $\lambda$ about the high-field limit with the long-range Ising interactions acting as a perturbation to 
\begin{equation}
  \mathcal{H}_0 =-\frac{1}{2}\sum_{\vec j}\sigma_{\vec j}^z~. \label{eq:H_0_tfim}
\end{equation}
The ground state of $\mathcal{H}_0$ is given by \(\ket{\uparrow\uparrow\cdots\uparrow}\) while the lowest excitations are single local spin flips. To obtain a quasi-particle (qp) description we perform a Matsubara-Matsuda transformation $\sigma_{\vec j}^x = \hat b_{\vec j}^\dagger+\hat b_{\vec j}^{\phantom{\dagger}}$ and \mbox{$\sigma_{\vec j}^z=1-2\hat{n}_{\vec j}$} \cite{Matsubara1956}. Here $\hat b_{\vec j}^{(\dagger)}$ are hardcore-boson annihilation (creation) operators and \mbox{$\hat{n}_{\vec j}\equiv \hat b^\dagger_{\vec j} \hat b^{\phantom{\dagger}}_{\vec j}$} counts the number of particles on site~$\vec j$. This gives \eqref{eq:H_tfim_orig} in a qp language
\begin{equation}
  \mathcal{H} =\sum_{\vec{j}} \hat{n}_{\vec j} -\lambda\sum_{\vec i\neq\vec j} \frac1{|\vec i-\vec j|^\alpha}\left( \hat b^\dagger_{\vec i} \hat b^\dagger_{\vec j} + \hat b^\dagger_{\vec i} \hat b^{\phantom{\dagger}}_{\vec j} + {\rm H.c.}\right) \, , \label{eq:H_tfim_orig_boson}
\end{equation}
up to a constant of \(-N/2\).

Next we apply perturbative continuous unitary transformations (pCUTs) \cite{Knetter2000} using white graphs \cite{Coester2015} to transform Eq.~\eqref{eq:H_tfim_orig}, order by order in $\lambda$, to an effective qp-conserving Hamiltonian $\mathcal{H}_{\rm eff}$ as it has been done successfully for the one-dimensional lrTFIM \cite{Fey2016}. As a consequence, $\mathcal{H}_{\rm eff}$ is block-diagonal in the qp number \mbox{\( \hat{Q}\equiv\sum_\vec{i}\hat{n}_\vec{i} \) } and the quantum many-body system is mapped to an effective few-body problem. Here we consider the one-qp block which can be expressed as 
\begin{equation}
  \mathcal{H}_{\rm eff}^{\rm 1qp} = E_0 + \sum_{\vec i,\vec{\delta}} a_{\vec\delta} \left( \hat{b}^\dagger_{\vec i} \hat b^{\phantom{\dagger}}_{\vec i + \vec\delta}+{\rm H.c.}\right) \, , \label{eq:H_eff_1qp}
\end{equation}
with the ground-state energy $E_0$ and the hopping amplitudes $a_{\vec\delta}$. The one-qp Hamiltonian \eqref{eq:H_eff_1qp} is diagonalized by Fourier transformation yielding \mbox{$\mathcal{H}_{\rm eff}^{\rm 1qp} = E_0+\sum_{\vec k} \omega(\vec k)\,\hat{b}^\dagger_{\vec k} \hat b^{\phantom{\dagger}}_{\vec k}$}. In the following, we focus on the one-qp gap $\Delta$ which is the minimum of the one-qp dispersion \mbox{$\omega (\vec k )=a_{\vec 0}+\sum_{\vec\delta\neq \vec{0}} a_\delta \cos\left( \vec{k} \cdot \vec{\delta}\right) $}. The gap is located at $\vec k=\vec 0$ in the ferromagnetic cases and at $\vec k=(\pi,\pi)$ [$\vec k=\pm(2\pi/3,- 2\pi/3)$] for the antiferromagnetic lrTFIM on the square [triangular] lattice in the $\alpha$-ranges we have studied (see Fig.~\ref{fig:graph_embeddings} for the definition of basis vectors). 
%A generalization to other lattices with multiple sites per unit cell is %straightforward.  

The pCUT determines the hopping amplitudes $a_{\vec\delta}$ and therefore the one-qp dispersion $\omega (\vec k )$ as a high-order series expansion in $\lambda$ in the thermodynamic limit. This can be done most efficiently via a full graph decomposition in linked graphs $\mathcal{G}$ exploiting the linked-cluster theorem \cite{Coester2015}. While for Hamiltonians with short-range interactions the main challenge lies in the generation of and calculation on linked graphs contributing in a given order, for long-range interactions the difficulty is shifted to the final embedding \cite{Fey2016}. In order $k$ perturbation theory all linked graphs with up to $k$ links may contribute in the calculation. The embedding of the graph-specific contribution $a_{\vec\delta}^{\mathcal{G}}$ to the hopping amplitudes $a_{\vec\delta}$ in Eq.~\eqref{eq:H_eff_1qp} in the infinite lattice requires the embedding of every single link infinitely many times due to the long-range character of the interaction. As a result of the infinite number of possible embeddings on the lattice for each graph, a conventional linked-cluster expansion becomes problematic. At this point white graphs are essential \cite{Coester2015}, since they allow to extract the generic linked contributions from graphs in a first step while the embedding on a specific lattice is done only at the end of the calculation. Here we perform the pCUT on each graph by introducing different couplings $\lambda_j^{\mathcal{G}}$ with $j\in\{1,\ldots, n\}$ on the $n$ links of $\mathcal{G}$. The resulting linked contributions in terms of the $\lambda_j^{\mathcal{G}}$ are then embedded in the infinite system by identifying the sites of graph $\mathcal{G}$ with the sites of the lattice and therefore replacing $\lambda_j^{\mathcal{G}}$ with the true interactions $\lambda\, |\bf i -\bf j|^{-\alpha}$ for each pair of sites $\bf i$ and $\bf j$ on the lattice. 

%\begin{figure}[thb]
%	\centering
%	\includegraphics[width=.6\columnwidth]{_graph_6_embeddings2.pdf}
%	\caption{Various embedding examples of the three-site graph on the square lattice for a fixed $\vec\delta$.}
%	\label{fig:graph_embeddings}
%\end{figure}

Let us illustrate the white graph expansion by considering the hopping amplitude $a_{\vec{\delta}}^{\mathcal{G}}$ with $\vec{\delta}= \vec{i_2}-\vec{i_0}$ in second-order perturbation theory on a linked three-site chain graph with sites $\vec{i_0}$, $\vec{i_1}$, and $\vec{i_2}$ (see Fig.~\ref{fig:graph_embeddings}). We introduce the two coupling constants $\lambda_1^{\mathcal{G}}$ and $\lambda_2^{\mathcal{G}}$ on this graph and obtain the generic pCUT contribution \(-\frac12\lambda_0^{\mathcal{G}}\lambda_1^{\mathcal{G}}\) to the hopping amplitude $a_{\vec{\delta}}^{\mathcal{G}}$. Embedding this term on the lattice in the thermodynamic limit, every hopping amplitude $a_{\vec{\delta}}$ gets infinitely many contributions: the $\vec{\delta}$ is set by fixing the two sites $\vec{i_0}$ and $\vec{i_2}$, but the remaining site $\vec{i_1}$ can be placed on any other site of the lattice as illustrated in Fig.~\ref{fig:graph_embeddings}. After Fourier transformation we get the following contribution of this graph to the one-qp dispersion $\omega (\vec k)$
\begin{equation}\label{eq:nested_sums}
         -\frac{\lambda^2}{2} \sum_{\vec{i_1}\neq \vec{0}} \sum_{\substack{\vec{\delta}\neq \vec{0} \\ \vec{\delta}\neq \vec{i_1}}} \frac1{|\vec{i_1}|^{\alpha}}\frac1{|\vec{i_1}-\vec{\delta}|^{\alpha}}\cos(\vec{k}\cdot\vec{\delta})~.
\end{equation}
In general, the embedding procedure leads to the occurrence of nested infinite sums like \eqref{eq:nested_sums}. In the most complex nested sum in order $k$ there are $d\,k$ infinite sums, where $d$ is the dimension of the lattice. Here we calculated series expansions of order $9$ for $\Delta$, which results in $18$ nested sums for the most difficult terms. In total, a number of $1068$ different graphs have to be treated for each $\vec k$ and $\alpha$. Let us stress that these calculations are tremendously more demanding compared to those in the one-dimensional lrTFIM where a brute-force evaluation of the nested sums up to order 8 is still feasible \cite{Fey2016}. In two dimensions, an analogue calculation would only reach order 4, which is certainly not sufficient to extract quantum-critical properties of the lrTFIM. 
Substantial progress is therefore needed to reach order 9 which we achieved by implementing classical Markov-Chain Monte Carlo (MCMC) integration techniques. The infinitely-large configuration space of embeddings is sampled in order to calculate the coefficients $c_k$ of the gap series $\Delta=\sum_{k=1}^9 c_k\,\lambda^k$.
Details on the implementation and the performance of the MCMC are given in the Supplementary Materials \cite{suppl18}. For a given lattice, exponent $\alpha$, and momentum $\vec k$, we sort the resulting nested sums of all graphs in a given perturbative order by the number of sites $N_{\mathcal{G}}$ of the graphs. Then a separate MCMC calculation is performed for each $N_{\mathcal{G}}$ in every order from 1 to 9. Effectively, the problem is reduced to the computation of the classical partition function of an $N_{\mathcal{G}}$-mer with many-body interactions with respect to a linear molecule. Joining all contributions from the various MCMC calculations, we obtain numerical estimates for the coefficients $c_k$ which are given in \cite{suppl18}. Most importantly, the numerical uncertainty is small enough in the $c_k$ so that any conclusion drawn below is not affected. 

The final series of the gap have to be extrapolated in order to extract quantum-critical properties of the \mbox{lrTFIM}. As for the one-dimensional lrTFIM \cite{Fey2016}, we expect second-order quantum phase transitions out of the high-field quantum paramagnet so that the one-particle gap $\Delta$ closes as $(\lambda-\lambda_{\rm c})^{z\nu}$ near the quantum critical point $\lambda_{\rm c}$. Here $z$ is the dynamical and $\nu$ the correlation-length critical exponent. The quantities $\lambda_{\rm c}$ and $z\nu$ are then estimated by DlogPad\'e extrapolation of the gap series. As error bars for these quantities we use the standard deviation of non-defective DlogPad\'e extrapolants. Further details of the extrapolation and our error estimates are given in \cite{suppl18}.

% Application
%%%%%%%%%%%%%%%%%%%%%%%%%%%%%%%%%%%%%%%%%%%%%%%%%%%%%%%%%%%%%%%%%%%%%%%%%%%%%%%%%%%%%%%%%%%%
{\it{Results:}}
We apply our approach to the lrTFIM on the square and triangular lattice, both for a ferromagnetic and an antiferromagnetic Ising exchange. The main goal is to determine the quantum phase diagram and to analyze the universality classes as a function of $\alpha$. 
%We present data for a ferromagnetic and an antiferromagnetic interaction for a %square and a triangular lattice.
%The universality class for the unfrustrated systems with a nearest-neighbor %interaction (both ferromagnets and the square antiferromagnet) is (2+1)-Ising %{\color{red} \cite{Pfeuty1971,Triang.ferro?}}. The antiferromagnetic nearest-%neighbor Ising model on the triangular lattice is frustrated and suspected to be %in the 3D-XY universality class\cite{Moessner2003,Powalski2013}.

%Figure N: Ferromagnetic critical value and exponent
%%%%%%%%%%%%%%%%%%%%%%%%%%%%%%%%%%%%%%%%%%%%%%%%%%%%%%%%%%%%%%%%%%%%%%%%%%%%%%%%%%%%%%%%%%%%%%%%%%%%%%%%%%%%%%%%%%%%%%%%%%%%%%%
\begin{figure} [t!]
 \includegraphics[width=\columnwidth]{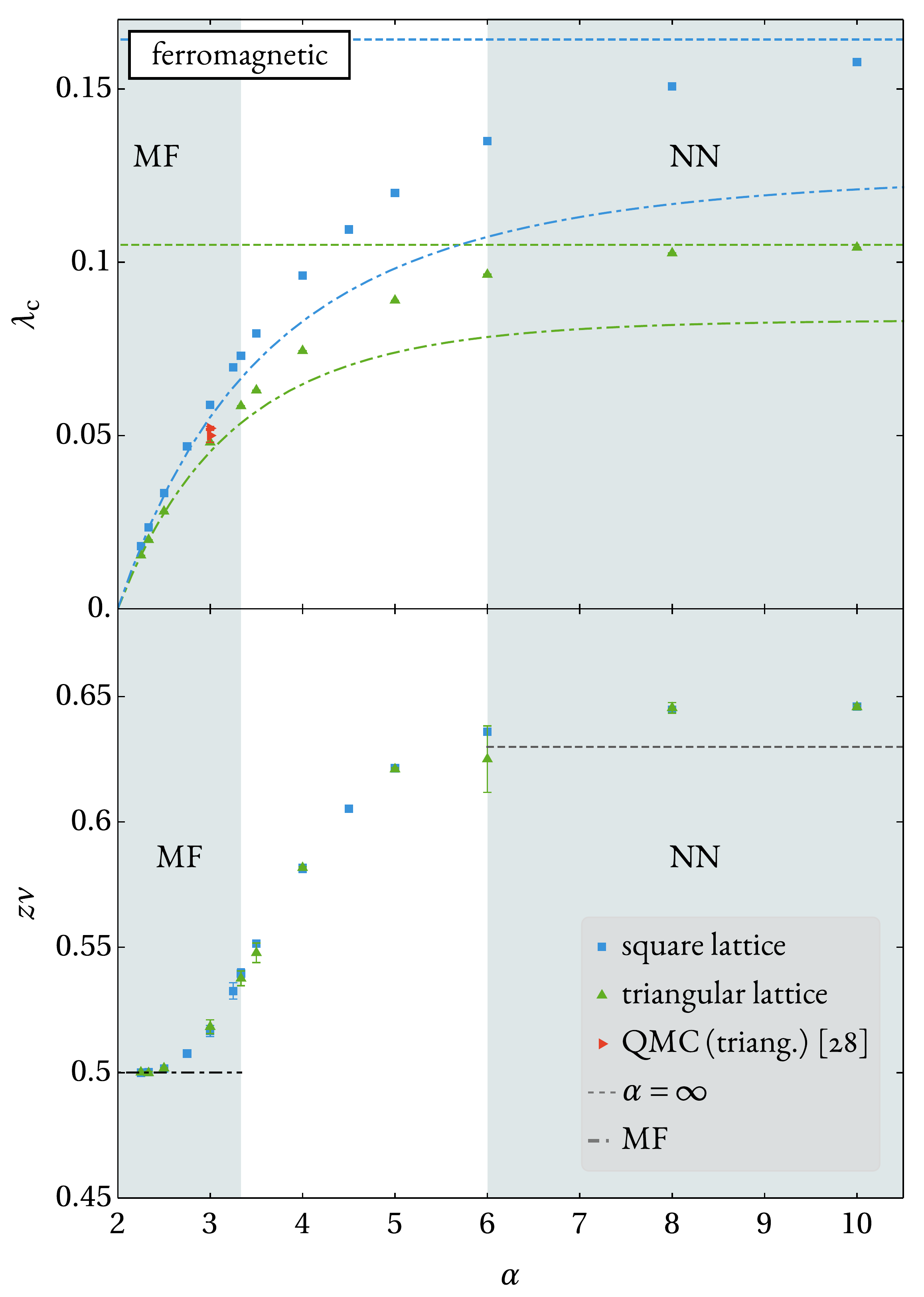}
 \caption{Critical point \(\lambda_{\mathrm{c}}\) (upper panel) and exponent \(z\nu\) (lower panel) are shown as squares (triangles) for the ferromagnetic lrTFIM on the square (triangular) lattice. Error bars represent the standard deviation of non-defective DlogPad\'e extrapolants. Shaded areas correspond to MF (left) and nearest-neighbor (NN) (right) universality. ({\it Upper panel}) Dashed lines indicate the quantum critical points \(\lambda_{\mathrm{c}}=0.16421 \) \cite{Hamer2000} (square lattice) and \(\lambda_{\mathrm{c}}=0.105\) \cite{Yanase1977} (triangular lattice) for the nearest-neighbor TFIM. MF results are given as dot-dashed lines and the quantum Monte Carlo data as red triangles (see \cite{Humeniuk2016}). ({\it Lower panel}) The upper (lower) dashed line refers to \(z\nu\approx 0.63\) \cite{Kos2016} (\( z\nu=0.5\)) of the nearest-neighbor TFIM (in MF).}
 \label{fig:square_triangular__ferro__values_exponents}
\end{figure}
%%%%%%%%%%%%%%%%%%%%%%%%%%%%%%%%%%%%%%%%%%%%%%%%%%%%%%%%%%%%%%%%%%%%%%%%%%%%%%%%%%%%%%%%%%%%%%%%%%%%%%%%%%%%%%%%%%%%%%%%%%%%%%%

{\it{Ferromagnetic interaction:}}
In this case the lrTFIM is in the 3D-Ising universality class for \mbox{\(\alpha\rightarrow\infty\)} on both lattices with a critical exponent \mbox{\(z\nu\approx 0.63\)} \cite{Kos2016}. As a function of $\alpha$, a similar behavior as for the 1D lrTFIM is expected \cite{Fey2016,Dutta2001}, where the critical exponent $z\nu$ varies continuously in a certain range of $\alpha$ from 2D-Ising to the MF value \(z\nu=0.5\). However, the boundaries in $\alpha$ of continuously varying exponents are shifted to $\alpha=10/3$ and $\alpha=6$ \cite{Dutta2001}. In Fig.~\ref{fig:square_triangular__ferro__values_exponents} we show our results for \(\lambda_{\mathrm{c}}\) and \(z\nu\) for both lattices (green and blue squares and triangles). We also display MF results as in Ref.~\cite{Humeniuk2016} (dot-dashed lines) and the quantum Monte Carlo data point for $\alpha=3$ on the triangular lattice (red triangles) \cite{Humeniuk2016}, which agrees well with our data. For a large \(\alpha=10\) the critical value $\lambda_{\mathrm{c}}$ is already very close to its nearest-neighbor correspondent. Strengthening the longer-range couplings by reducing \(\alpha\) stabilizes the \(\mathcal{Z}_2\)-broken phase and \(\lambda_{\mathrm{c}}\) decreases. In the limit \(\alpha\rightarrow2\) the phase transition happens at \(\lambda_{\mathrm{c}}\rightarrow0\), while for exactly \(\alpha=2\) the sums diverge and \eqref{eq:H_tfim_orig} becomes ill-defined. However, we stress that our results agree with MF calculations (dot-dashed lines in Fig.~\ref{fig:square_triangular__ferro__values_exponents}) even in the regime $\alpha\leq 2.5$, where the MF ansatz is expected to be quantitatively correct. 

Next we discuss the behavior of $z\nu$. It is known that the DlogPad\'{e} extrapolation slightly overestimates critical exponents, since it ignores subleading corrections to the critical behavior. As a consequence, for both lattices, the estimate $z\nu\approx 0.65$ for large $\alpha$ is about 3\% too large compared to the known value $z\nu\approx 0.63$ \cite{Kos2016} of the nearest-neighbor TFIM \cite{Kos2016,Oitmaa1991}. In the opposite limit of small $\alpha$ the critical exponent $z\nu$ approaches the MF value $0.5$ confirming the expected MF limit. In between we find an interesting continuous variation of $z\nu$ from the MF value to that of the 3D-Ising universality class. Note that we attribute the deviations from $0.5$ for $\alpha\leq 10/3$ to limitations of the extrapolation which neglects the subleading multiplicative logarithmic correction $p$ at $\alpha=10/3$ (for a definition of $p$ see \cite{suppl18}). Indeed, when extracting $p$ for $\alpha=10/3$ from the DlogPad\'e extrapolation by fixing $\lambda_{\rm c}$ and $z\nu=1/2$ as for the one-dimensional \mbox{lrTFIM} \cite{Fey2016}, we find $p=-0.17(4)$ ($p=-0.143(7)$) for the square (triangular) lattice. These values are remarkably close to $p=-1/6$ which is the prediction for the 3d TFIM from perturbative RG and series expansions \cite{Larkin1969,Brezin1973,Wegner1973,Weihong1994,Coester2016}. The quantum-critical behavior induced by the long-range Ising interaction can therefore effectively be understood in terms of the nearest-neighbor TFIM in an effective spatial dimension $d_{\rm eff}$. Furthermore, we stress that the estimated critical exponents agree extremely well on both lattices. This property can be seen as a kind of meta-universality because the universality class of both models changes identically with the parameter~\(\alpha\).

{\it{Antiferromagnetic interaction:}}
Here we expect an inherently different behavior not only with respect to the ferromagnetic case but also when comparing both lattices. Already in the nearest-neighbor limit $\alpha\rightarrow\infty$ one finds two different universality classes, since the TFIM on the triangular lattice displays 3D-XY universality due to the strong geometric frustration. On the square lattice, the long-range Ising interaction introduces also frustration which is, however, expected to be weaker. For both lattices there is no MF limit for small values of $\alpha$ and it is therefore not at all obvious how the quantum critical behavior changes as a function of $\alpha$ in these frustrated systems. 

%Figure N: Antiferromagnetic critical value and exponent
%%%%%%%%%%%%%%%%%%%%%%%%%%%%%%%%%%%%%%%%%%%%%%%%%%%%%%%%%%%%%%%%%%%%%%%%%%%%%%%%%%%%%%%%%%%%%%%%%%%%%%%%%%%%%%%%%%%%%%%%%%%%%%%
\begin{figure} [t!]
 \includegraphics[width=\columnwidth]{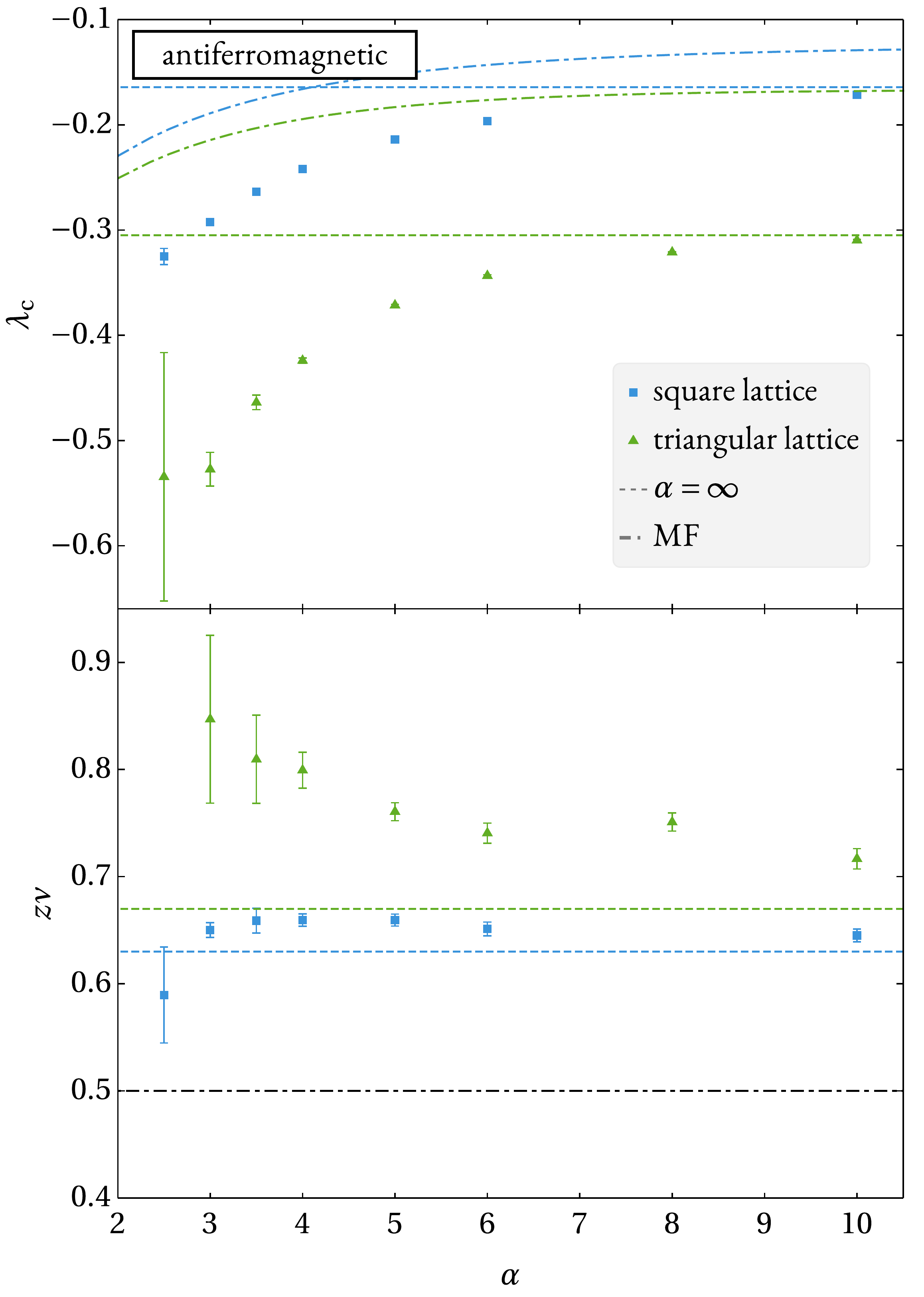}
 \caption{Critical point \(\lambda_{\mathrm{c}}\) (upper panel) and exponent \(z\nu\) (lower panel) are shown as squares (triangles) for the antiferromagnetic lrTFIM on the square (triangular) lattice. Error bars indicate the standard deviation of non-defective DlogPad\'e extrapolants. ({\it Upper panel}) Dashed lines correspond to \mbox{\(\lambda_{\mathrm{c}}=-0.16421(1)\)} \cite{Hamer2000} (\(\lambda_{\mathrm{c}}=-0.305\)\cite{Moessner2003,Powalski2013}) for the TFIM on the square (triangular) lattice. ({\it Lower panel}) Dashed lines refer to 3D-Ising exponent \( z\nu\approx 0.63\) \cite{Kos2016} and 3D-XY exponent \( z\nu\approx 0.67\) \cite{Gottlob1994}.} 
 \label{fig:square_triangular__antiferro__values_exponents}
\end{figure}
%%%%%%%%%%%%%%%%%%%%%%%%%%%%%%%%%%%%%%%%%%%%%%%%%%%%%%%%%%%%%%%%%%%%%%%%%%%%%%%%%%%%%%%%%%%%%%%%%%%%%%%%%%%%%%%%%%%%%%%%%%%%%%%

Our results for $\lambda_{\mathrm{c}}$ and $z\nu$ are shown for both lattices in Fig.~\ref{fig:square_triangular__antiferro__values_exponents}. As expected, stronger competing interactions introduced by decreasing $\alpha$ stabilize the quantum paramagnet. We observe that the MCMC becomes less reliable for $\alpha$ close to 2. Furthermore, small $\alpha$ values lead to alternating series in $|\lambda|$ with extremely large coefficients $c_k$ which are hard to extrapolate (see also \cite{suppl18}). This results in rather large error bars for $\alpha\leq 3$ as can be seen in Fig.~\ref{fig:square_triangular__antiferro__values_exponents}. Consequently we only show results for \(\alpha\geq 2.5\). \footnote{Note that the critical point for \(\alpha=3\) from quantum Monte Carlo simulations \cite{Humeniuk2016} lies above our results. However, this data point has been produced by a data collapse assuming MF exponents which we think is not appropriate with respect to our findings.}

As outlined above, limitations in the extrapolation lead to a slightly overestimated $z\nu$ for large $\alpha$ \cite{Powalski2013,Oitmaa1991}. 
Decreasing $\alpha$, $z\nu$ stays almost constant and close to the value of the nearest-neighbor TFIM which is different for the two lattices. On the square lattice, only for $\alpha=2.5$ it is below the $\alpha\rightarrow\infty$ limit but with a significantly larger uncertainty in the extrapolation. On the triangular lattice, we have larger uncertainties in the estimates of $\lambda_{\mathrm{c}}$ and $z\nu$, which is likely a consequence of the stronger frustration. We find a modest increase in $z\nu$ for decreasing $\alpha$, but it is still within the error bars of our data that it stays constant in the full range of displayed $\alpha$. Our results are therefore compatible with the scenario that both frustrated systems remain in the universality class of the nearest-neighbor TFIM independent of $\alpha$ in a similar fashion as deduced for the one-dimensional lrTFIM \cite{Sun2017}. 

{\it{Conclusions:}}
We investigated the largely unexplored interplay of long-range interactions, quantum fluctuations, and frustration in 2d quantum magnets directly in the thermodynamic limit. This was achieved by a technical breakthrough combining high-order series expansions with classical Monte Carlo simulations.  

{\it{Acknowledgments:}}
During the completion of this work, we became aware of the numerical work by Saadatmand et al.~\cite{Saadatmand2018} who studied the antiferromagnetic triangular lattice lrTFIM on infinitely long cylinders. We gratefully acknowledge the compute resources and support provided by the HPC group of the Erlangen Regional Computing Center (RRZE).

\clearpage

\renewcommand{\theequation}{S\arabic{equation}}
\renewcommand{\thefigure}{S\arabic{figure}}
\renewcommand{\citenumfont}[1]{S#1}

\newcommand\FIXME[1]{{\bf FIXME \{}#1{\bf \}}}

%%%%%%%%%% Merge with supplemental materials %%%%%%%%%%

\widetext
%
%%%%%%%%%%%%%%%%%%%%%%%
%%%%%%%%%%%%%%%%%%%%%%%
\begin{center}
\textbf{\large Supplementary Materials to "Quantum criticality of two-dimensional quantum magnets with long-range interactions".}

S. Fey, S.C. Kapfer, and K.P. Schmidt.
\end{center}
%%%%%%%%%%%%%%%%%%%%%%%
%%%%%%%%%%%%%%%%%%%%%%%
%%%%%%%%%%%%%%%%%%%%%%%

\setcounter{equation}{0}
\setcounter{figure}{0}
\setcounter{table}{0}
\setcounter{page}{1}
\makeatletter

\thispagestyle{empty}

This Supplementary Material contains a detailed description of the Monte Carlo integration used to evaluate the nested infinite sums, tables of the series coefficients of all gap series, and a description of the extrapolation scheme applied to the gap series.  

%====================================================================================================================
% Evaluation of high-dimensional sums using Monte-Carlo integration
%====================================================================================================================
\section{Evaluation of nested infinite sums using Monte Carlo integration}
\label{sec:MC}

For a calculation of critical parameters the series coefficients, which are given as high-dimensional infinite sums, must be evaluated numerically. The number of summations within a nested sum grows as $D=d\,k$ with the expansion order $k$ and the lattice dimension \(d\). With a maximum order of the series expansion of \(k_{\mathrm{max}}=9\) on two-dimensional lattices we end up with $D_{\mathrm{max}}=18$ for the two-dimensional square and triangular lattice treated in the present paper. It is well-known that Monte Carlo integration is well-suited to the evaluation of high-dimensional sums as the asymptotic error only grows as \(C'/\sqrt{\mathcal{N}_{\text{steps}}}\), with the total number of Monte-Carlo steps \(\mathcal{N}_{\text{steps}}\) \cite{Krauth2006}.
There remains, however, an indirect dependency on the dimensionality, as the
constant $C'$ depends on the nature of the problem.

The final goal of the MC integration is the computation of the nested sum
\begin{align}
    I[f]
&:=
    \sum_{a} f(a)
\intertext{where a configuration $a\equiv\{\vec{i_\nu}\}$ comprises concrete
integral coordinates for all positions of the sites $\vec{i_\nu}$, i.\,e., a
geometric embedding of all graphs with $N_\mathcal{G}$ sites into the physical lattice, and $f(a)$
is the target integrand,}
    f(a)
&:= \sum_{\bar{\mathcal{G}}} \sum_{\mu=1}^{N_\mathcal{G}} \sum_{\nu=1}^{\mu} C^{\,\bar{\mathcal{G}}}_{\mu,\nu} \left( \prod_{\vec{\ell}} g^{n_{\vec{\ell}}} ( {\vec{\ell}})\right)  \cos[\vec{k}\cdot (\vec{i}_\mu-\vec{i}_\nu)]~,
\end{align}
where the first sum runs over all graphs $\bar{\mathcal{G}}$ with $N_\mathcal{G}$ sites and $C^{\,\bar{\mathcal{G}}}_{\mu,\nu}$ represents the graph-specific pCUT contribution for a hopping between sites $\vec{i}_\nu$ and $\vec{i}_\mu$. The product $\prod_{\vec{\ell}}$ is taken over all links $\vec{\ell}\equiv \vec{i}_\tau-\vec{i}_\xi$ with pairs of sites $\vec{i}_\tau$ and $\vec{i}_\xi$ of graph $\bar{\mathcal{G}}$ and $g^{n_{\vec{\ell}}} ( {\vec{\ell}})\equiv 1/|\vec{i}_\tau-\vec{i}_\xi|^{n_{\vec{\ell}}\alpha}$ with $n_{\vec{\ell}}\in\mathbb{N}$. Note that we set $f(a)\equiv 0$ whenever two graph sites are embedded on the same lattice site.

To evaluate the infinite sums, we use importance sampling with respect to some
probability weight $\pi(a)$,
\begin{equation}
    I[f] = \sum_a \frac{\pi(a)}{Z} \frac{Z}{\pi(a)} f(a) = Z \left\langle \frac{f(a)}{\pi(a)}~\right\rangle_{\pi}~\label{eq:MC_integral_f},
\end{equation}
where the angular brackets \(\langle \cdot \rangle_\pi\) denote the average with
respect to the probabilities \(\pi(a)/Z\), and $Z=\sum_a \pi(a)$ is
the associated partition function.
The partition function 
$Z$ cannot be computed directly in MC, but may be eliminated by evaluating an
analytically tractable reference integrand $f_0$ along with the target integrand
\begin{align}
    I[f] = I[f_0] \times \left.
        \left\langle \frac{f(a)}{\pi(a)}~\right\rangle_{\pi}
    \middle/
        \left\langle \frac{f_0(a)}{\pi(a)}~\right\rangle_{\pi}
        \right.
\end{align}
for which we use
\begin{equation}
	f_0(a) =  \prod_{\nu=1}^{N_{\mathcal{G}}-1}\frac{1}{(1+\Delta i_{\nu,\mathrm 1})^\rho}\frac{1}{(1+\Delta i_{\nu,\mathrm 2})^\rho}~,
\end{equation}
with \(\Delta i_{\nu,\kappa}=|i_{\nu+1,\kappa}-i_{\nu,\kappa}|\), \(\kappa\in\{\mathrm 1,\mathrm 2\}\) and the number of graph sites $N_{\mathcal{G}}$.
Note that this sum contains assignments of the site coordinates that
which violate the hard-core constraint in $f$.
While the exponent \(\rho\) is an arbitrary parameter (it has to be larger than one for the sum to converge), it is useful to choose \(\rho=\alpha/2\) to obtain the same asymptotics as for the target integrand. For large values of $\alpha$ a better convergence is obtained for \(\rho<\alpha/2\), so we choose $\rho=3$ for $7\leq\alpha<9$ and $\rho=3.5$ for $9\leq\alpha$. The value of the reference sum is easily calculated
\begin{align}
	I[f_0] &= \sum_{\substack{\Delta i_{1,\mathrm 1} = -\infty \\ \Delta i_{1,\mathrm 2} = -\infty }}^\infty \dots \sum_{\substack{\Delta i_{N_{\mathcal{G}}-1,\mathrm 1} = -\infty \\ \Delta i_{N_{\mathcal{G}}-1,\mathrm 2} = -\infty }}^\infty 
		\frac{1}{(1+\Delta i_{1,\mathrm 1})^\rho}\frac{1}{(1+\Delta i_{1,\mathrm 2})^\rho} \dots \frac{1}{(1+\Delta i_{N_{\mathcal{G}}-1,\mathrm 1})^\rho}\frac{1}{(1+\Delta i_{N_{\mathcal{G}}-1,\mathrm 2})^\rho} \notag \\
		&= \left( 2\zeta(\rho) - 1 \right) ^ {2(N_{\mathcal{G}}-1)}~.
\end{align}
For optimal convergence, we use the probability weights as
\begin{equation}
	\pi(a)=\sqrt{C^2 {f_0}^2(a) + f_{\phantom{0}}^2(a) } ~,
\label{defTheWeights}
\end{equation}
with a constant $C$ that is numerically determined such that both summands have an equal magnitude. This constant is found automatically in a brief in-advance calibration run of the Monte Carlo integration by comparing the values of \(f(a)\) and \(f_0(a)\).
(We use a fixed seed for the random number generator in all calibrations.)
After the determination of $C$ the actual calculation can be done, with independent
random number streams for each process.

\begin{figure} [t!]
 \includegraphics[width=.4\columnwidth]{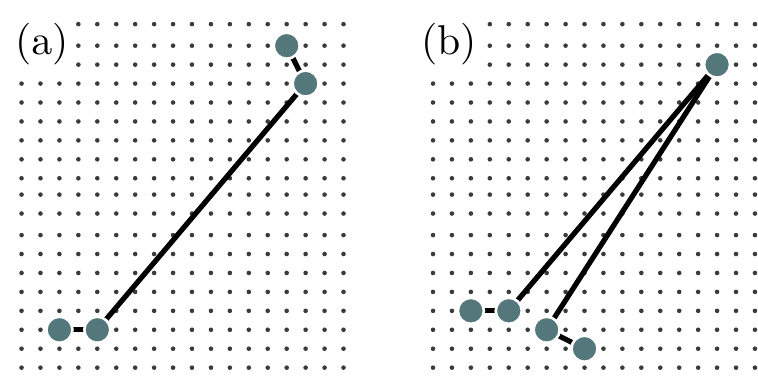}
 \caption{Two states with a low weight and therefore a low contribution to the integral. If the system is in one of these states, moves of type 2 for (a) and type~3 for (b) are required for the system to relax quickly to a configuration with a larger weight while still respecting the detailed balance condition.}
 \label{fig:MC_annoying_configurations}
\end{figure}

To sample the probability distribution Eq.~\eqref{defTheWeights}, we employ
Markov-chain Monte Carlo (MCMC), more specifically the Metropolis-Hastings
algorithm \cite{Hastings1970}.
In each step, a new configuration based on the current state is proposed according to the following scheme:
\begin{enumerate}
	\item With probability $0.7$, a randomly-chosen site $\vec i_\nu$ is shifted by a random vector \(\vec{\Delta} = (\Delta_1, \Delta_2)\), with $\Delta_\kappa$, $\kappa\in\{\mathrm 1,\mathrm 2\}$, and $\Delta_\kappa$ uniformly chosen in the range $\{-N_{\mathcal{G}},-N_{\mathcal{G}}+1, \dotsc,N_{\mathcal{G}}\}$.
	\item With probability $0.2$, two sites $\vec{i_\nu}$ and $\vec{i_{\nu+1}}$ are randomly selected and a new distance is randomly chosen. Both values $\Delta_\alpha$ for the new distance \(\vec{\Delta}\) are drawn from a $\zeta$-distribution with exponent $\rho$. Then, all sites with index $\geq\nu+1$ are shifted by \(\vec{\Delta}-(\vec{i_{\nu+1}}-\vec{i_{\nu}})\).
	\item With probability $0.1$ the same steps as before are taken. But here, not the whole chain with index $\geq\nu+1$ is moved. Instead only the single site with index $\nu$ is moved to the new position.
\end{enumerate}
The above steps ensure that the system can recover from configurations that have a very low probability of occurrence such as those shown in Fig. \ref{fig:MC_annoying_configurations}.

The proposed moves are accepted with the Metropolis-Hastings probability
\begin{equation}
	p_{\mathrm{acc}}(a\rightarrow b) = \mathop{\mathrm{min}}\left(1,\frac{\pi(\{\vec{i}_\nu^{b}\})\mathcal{A}(b\rightarrow a)}{\pi(\{\vec{i}_\nu^{a}\}) \mathcal{A}(a\rightarrow b)}\right),
\end{equation}
where for the cases 2 and 3
\begin{equation}
	% \mathcal{A}(a\rightarrow b) = \frac{\Delta_{a\rightarrow b,1}^{-\rho}\Delta_{a\rightarrow b,2}^{-\rho}}{\zeta(\rho)^2}
	\mathcal{A}(a\rightarrow b) = \frac{(1+|\Delta_{a\rightarrow b,1)}|)^{-\rho}(1+|\Delta_{a\rightarrow b,2)}|)^{-\rho}}{\left(2\zeta(\rho)-1\right)^2} 
\end{equation}

is the probability to propose configuration \(b\equiv\{\vec{i}_\nu^b\}\) when the system is currently in configuration \(a\equiv\{\vec{i}_\nu^a\}\). \(\Delta_{a\rightarrow b,\kappa}\) with \(\kappa\in \mathrm 1,\mathrm 2\) is the value of the $\kappa$-coordinate of the newly chosen distance between the selected sites.
For step 1 the new random positions are uniformly distributed such that in this case \(\mathcal{A}\) is always $1/(2N_{\mathcal{G}}+1)$.

\begin{figure} [t!]
 \includegraphics[width=0.6\columnwidth]{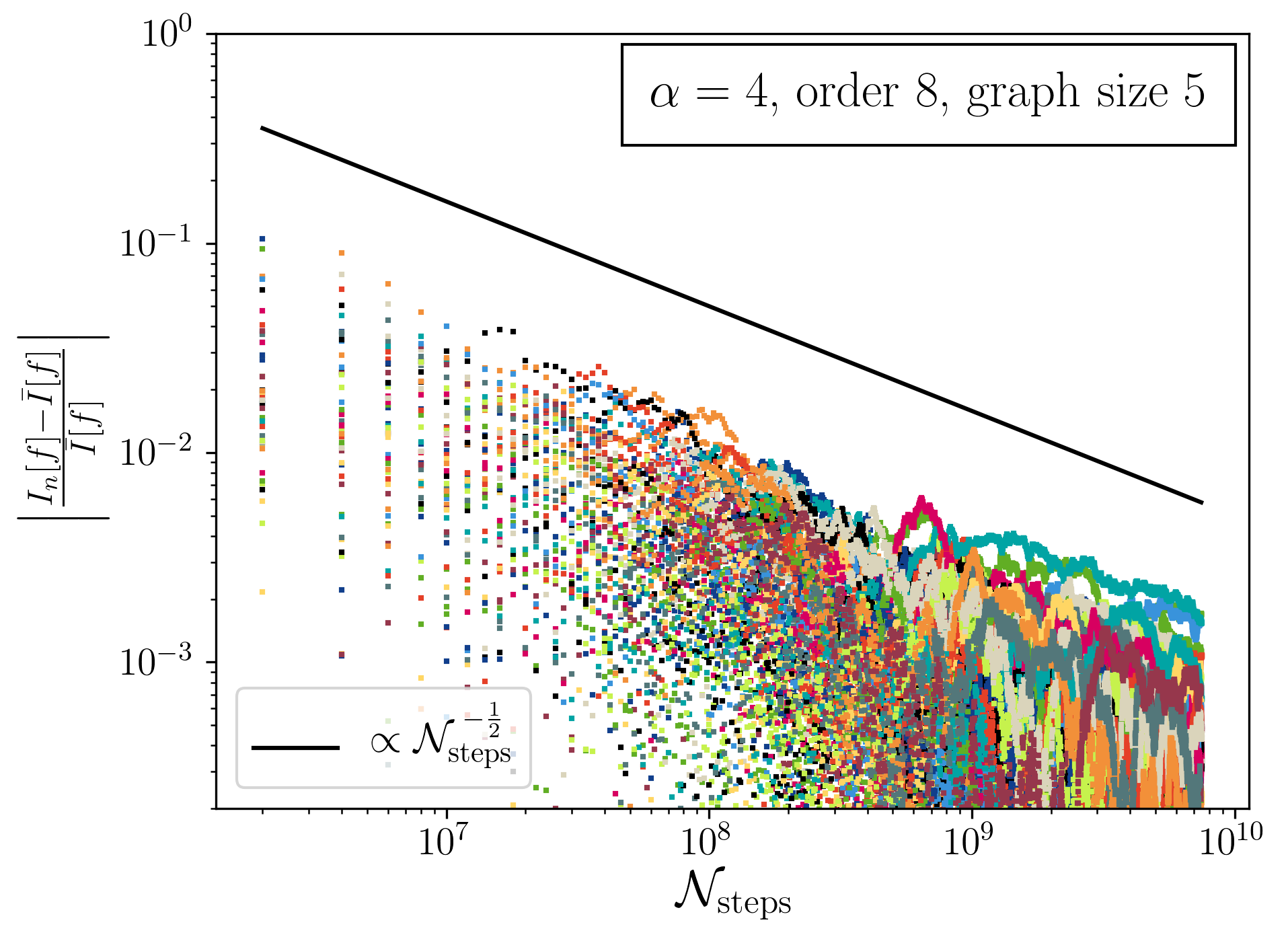}
 \caption{The deviation of the running mean $I_n[f]$ in the Monte-Carlo integration for 50 different seeds $n$ from the mean value of all 50 seeds combined \(\bar{I}[f]\). The chosen parameters are \(\alpha=4\) in order \(8\) for the sum of all possible five-vertex graphs. The Monte Carlo error displays the typical behavior of \(1/\sqrt{\mathcal{N}_{\mathrm{steps}}}\) where \(\mathcal{N}_{\mathrm{steps}}\) is the total number of MCMC steps.}
 \label{fig:MC_running_mean}
\end{figure}

As one representative example, the convergence for a calculation for all five-vertex graphs in order 8 is shown in Fig. \ref{fig:MC_running_mean} for $\alpha=4$. It can be clearly seen that the error of the calculation as asymptotically given as an inverse square root of the total number of steps \(\mathcal{N}_{\mathrm{steps}}\).

%Figure N: Monte-Carlo deviation behavior
%%%%%%%%%%%%%%%%%%%%%%%%%%%%%%%%%%%%%%%%%%%%%%%%%%%%%%%%%%%%%%%%%%%%%%%%%%%%%%%%%%%%%%%%%%%%%%%%%%%%%%%%%%%%%%%%%%%%%%%%%%%%%%%

\begin{figure} [t!]
 \includegraphics[width=.4\columnwidth]{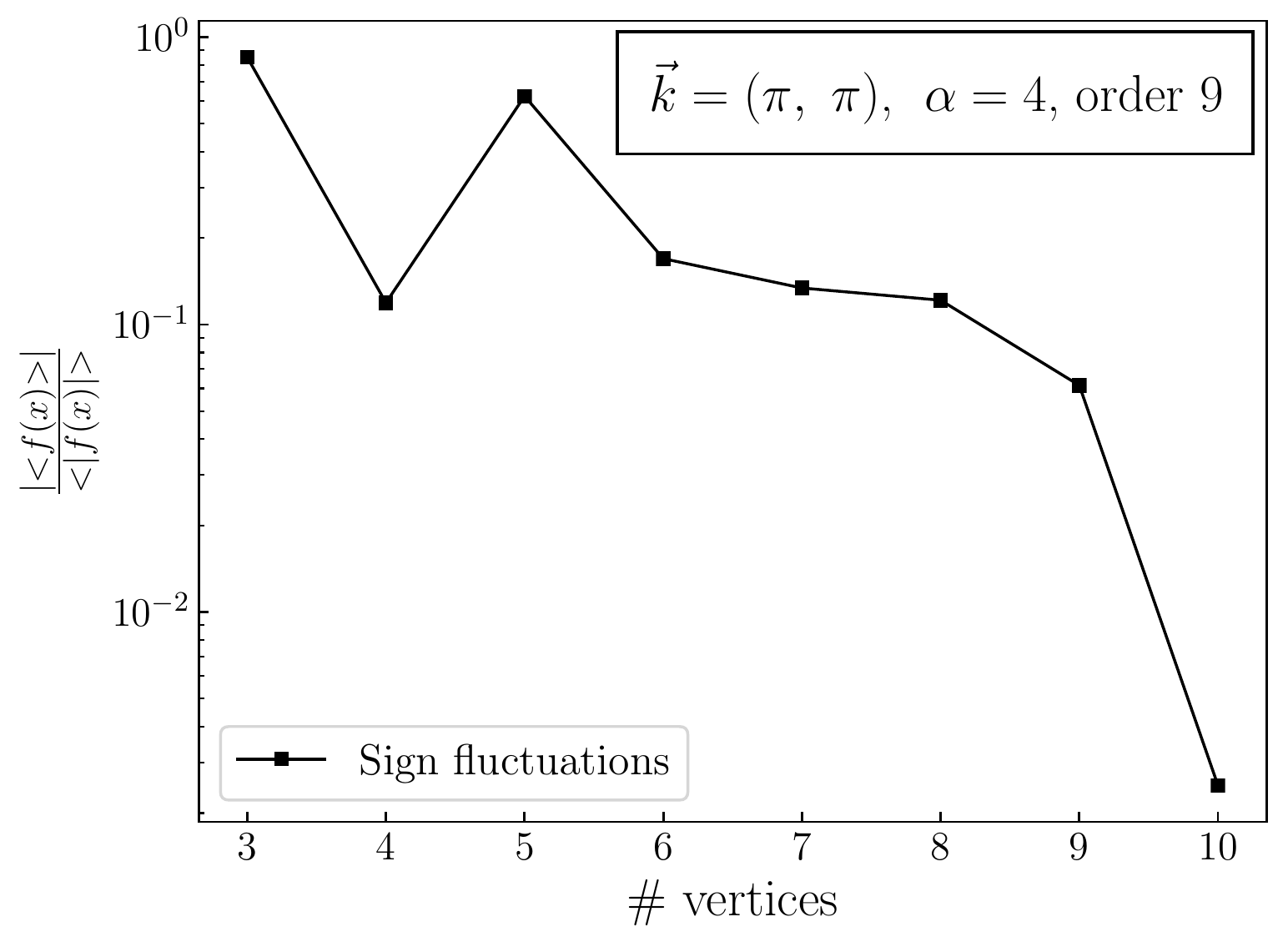}
 \caption{The dependence of the integrand sign on the graph size for an antiferromagnetic interaction with $\alpha=4$ in order $9$ on the square lattice (for the gap momentum $\vec k=(\pi,\pi)$).}
 \label{fig:MC_sign_behavior}
\end{figure}

Convergence of the MC integration is poor when the integral is much smaller
than the absolute integral, i.\,e., $I[f] \ll I[|f|]$. In the present case this is a problem especially for the antiferromagnetic interaction on both lattices. This can be seen most clearly for the square lattice where a hopping between two sites with distance $\vec{\Delta i_\nu}$ contributes terms proportional to \(\cos[(\pi,\pi)^T \cdot\vec{\Delta i_\nu}]\) to the one-particle gap at $\vec k=(\pi,\pi)$. However, we find that the sign fluctuations are sufficiently small for all treated $\alpha$ on both lattices to obtain reliable results, see the tables given in the next section.
As a representative example, we illustrate these sign fluctuations for the highest calculated order 9 and $\alpha=4$ for the square lattice in Fig.~\ref{fig:MC_sign_behavior}. One observes that even for the largest possible graphs in this order (10 sites) the sign fluctuation is small enough for the calculations to converge in a reasonable amount of time.

%Figure N: Sign-dependence antiferro
%%%%%%%%%%%%%%%%%%%%%%%%%%%%%%%%%%%%%%%%%%%%%%%%%%%%%%%%%%%%%%%%%%%%%%%%%%%%%%%%%%%%%%%%%%%%%%%%%%%%%%%%%%%%%%%%%%%%%%%%%%%%%%%

%====================================================================================================================
% List of Coefficients
%====================================================================================================================
\onecolumngrid
\section{List of Coefficients}

We show all series' coefficients of the gap for both lattices as well as ferro- and antiferromagnetic Ising interactions in tables \ref{tab:coeff_square_f}-\ref{tab:coeff_triangular_af}. They are sorted by their momentum (for the ferro- and antiferromagnetic gap) and the respective lattice.

For each of the parameter sets with momentum $\vec k$, parameter $\alpha$, and order $k$ MCMC calculations as described above are performed for the different possible graph sizes in order $k$. To obtain reasonable error estimates of the coefficients we seed our programs with up to 50 different numbers for the most difficult calculations (small $\alpha$). The mean is obtained by averaging over all seeds; the standard deviation defines the error given in round brackets in below tables.

For the first order coefficients an analytic expression can be found. The ferromagnetic first order term on the square and triangular lattice with momentum \(\vec k = \left(0,0\right)^T\) are respectively given as
\begin{align}
	c_1^{\square,\,\mathrm F}(\alpha) &= -4 \zeta\left(\frac\alpha2\right) L_{-4}\left( \frac\alpha2\right) \\
	c_1^{\triangle,\,\mathrm F}(\alpha) &= -6 \zeta\left(\frac\alpha2\right) L_{-3}\left( \frac\alpha2\right)~.
\end{align}
For the antiferromagnetic cases we obtain the first-order coefficients
\begin{align}
	c_1^{\square,\,\mathrm {AF}}(\alpha) &= 4 \left(1-2^{1-\frac{\alpha }{2}}\right) L_{-4}\left(\frac{\alpha }{2}\right) \zeta \left(\frac{\alpha }{2}\right)\quad\text{for }\vec k = \left(\pi,\pi\right)^T\\
	c_1^{\triangle,\,\mathrm {AF}}(\alpha) &= 3 \left(1-3^{1-\frac{\alpha }{2}}\right) L_{-3}\left(\frac{\alpha }{2}\right) \zeta \left(\frac{\alpha }{2}\right)\quad\text{for }\vec k = \left(\frac23\pi,-\frac23\pi\right)^T~.
\end{align}
Here $\zeta(s)$ is the Riemann $\zeta$-function and $L_m(s)$ are Dirichlet $L$-series.

\begingroup
\squeezetable
\begin{table}[tb]
	\caption{Coefficients $c_k$ of the gap $\Delta$ for each order $k$ in the ferromagnetic lrTFIM on the square lattice (\(\vec k = \left(0,0\right)^T\)).}
	\label{tab:coeff_square_f}
	\centering
	\bgroup
	\setlength\tabcolsep{.8em}
	% \begin{ruledtabular}
		% \newcolumntype{e}[1]{D{e}{\mathrm{e}}{#1} }
		% \end{tabular}
		\begin{tabular}{cSSSSS}
		\toprule
		\diagbox{$k$}{$\alpha$} & \multicolumn{1}{c}{2.25} & \multicolumn{1}{c}{$\frac73$} & \multicolumn{1}{c}{2.5} & \multicolumn{1}{c}{2.75} & \multicolumn{1}{c}{3} \\
		\midrule
		2 & -3.7995(4)e2 & -2.25666(20)e2 & -1.11013(7)e2 & -5.66521(27)e1 & -3.61442(18)e1 \\
		3 & -1.04348(12)e4 & -4.7596(4)e3 & -1.62566(12)e3 & -5.8076(4)e2 & -2.89181(18)e2 \\
		4 & -3.6061(7)e5 & -1.26838(22)e5 & -3.0358(4)e4 & -7.7049(8)e3 & -3.03941(28)e3 \\
		5 & -1.3956(5)e7 & -3.7863(10)e6 & -6.3570(11)e5 & -1.14801(14)e5 & -3.5900(5)e4 \\
		6 & -5.786(4)e8 & -1.2109(5)e8 & -1.4272(4)e7 & -1.8363(5)e6 & -4.5598(9)e5 \\
		7 & -2.512(4)e10 & -4.0559(26)e9 & -3.3558(14)e8 & -3.0791(10)e7 & -6.0755(15)e6 \\
		8 & -1.128(4)e12 & -1.4049(19)e11 & -8.161(5)e9 & -5.3411(22)e8 & -8.378(4)e7 \\
		9 & -5.19(4)e13 & -4.988(14)e12 & -2.0349(20)e11 & -9.503(7)e9 & -1.1856(9)e9 \\[1em]

		 & \multicolumn{1}{c}{3.25} & \multicolumn{1}{c}{$\frac{10}{3}$} & \multicolumn{1}{c}{3.5} & \multicolumn{1}{c}{4} & \multicolumn{1}{c}{4.5} \\
		\midrule
		2 & -2.59986(9)e1 & -2.37180(12)e1 & -2.01472(9)e1 & -1.38799(6)e1 & -1.06952(4)e1 \\
		3 & -1.72329(13)e2 & -1.49030(27)e2 & -1.15001(7)e2 & -6.3440(4)e1 & -4.19616(30)e1 \\
		4 & -1.52263(15)e3 & -1.2542(5)e3 & -8.8709(8)e2 & -4.0104(4)e2 & -2.31058(28)e2 \\
		5 & -1.51153(17)e4 & -1.1856(8)e4 & -7.6827(12)e3 & -2.8427(4)e3 & -1.42888(26)e3 \\
		6 & -1.6165(4)e5 & -1.2080(13)e5 & -7.1812(14)e4 & -2.1819(4)e4 & -9.5833(30)e3 \\
		7 & -1.8143(5)e6 & -1.2916(25)e6 & -7.0459(20)e5 & -1.7579(5)e5 & -6.7474(29)e4 \\
		8 & -2.1087(8)e7 & -1.432(4)e7 & -7.1632(26)e6 & -1.4691(8)e6 & -4.932(4)e5 \\
		9 & -2.5158(17)e8 & -1.628(8)e8 & -7.477(5)e7 & -1.2613(13)e7 & -3.705(5)e6 \\[1em]

		 & \multicolumn{1}{c}{5} & \multicolumn{1}{c}{6} & \multicolumn{1}{c}{8} & \multicolumn{1}{c}{10} \\
		\midrule
		2 & -8.8236(4)e0 & -6.7887(4)e0 & -5.1497(5)e0 & -4.5319(9)e0 \\
		3 & -3.11638(24)e1 & -2.12938(21)e1 & -1.51239(27)e1 & -1.3308(5)e1 \\
		4 & -1.55106(23)e2 & -9.2128(18)e1 & -5.5384(24)e1 & -4.441(4)e1 \\
		5 & -8.7312(20)e2 & -4.6560(17)e2 & -2.6300(21)e2 & -2.114(4)e2 \\
		6 & -5.3181(17)e3 & -2.4950(14)e3 & -1.2107(22)e3 & -8.86(5)e2 \\
		7 & -3.4081(20)e4 & -1.4285(15)e4 & -6.464(18)e3 & -4.77(6)e3 \\
		8 & -2.2651(19)e5 & -8.380(12)e4 & -3.267(19)e4 & -2.19(5)e4 \\
		9 & -1.550(4)e6 & -5.115(19)e5 & -1.86(8)e5 & -1.24(9)e5 \\

		\bottomrule
		\end{tabular}

	% \end{ruledtabular}
	\egroup
	\end{table}
	\endgroup

\begingroup
\squeezetable
\begin{table}[tb]
	\caption{Coefficients $c_k$ of the gap $\Delta$ for each order $k$ in the ferromagnetic lrTFIM on the triangular lattice (\(\vec k = \left(0,0\right)^T\)).}
	\label{tab:coeff_triangular_f}
	\centering
	\bgroup
	\setlength\tabcolsep{.8em}
	% \begin{ruledtabular}
		% \newcolumntype{e}[1]{D{e}{\mathrm{e}}{#1} }
		% \end{tabular}
		\begin{tabular}{cSSSSS}
		\toprule
		\diagbox{$k$}{$\alpha$} & \multicolumn{1}{c}{2.25} & \multicolumn{1}{c}{2.33333} & \multicolumn{1}{c}{2.5} & \multicolumn{1}{c}{3} & \multicolumn{1}{c}{3.33333} \\
		\midrule
		2 & -5.2241(4)e2 & -3.13485(22)e2 & -1.57409(12)e2 & -5.4501(5)e1 & -3.72541(19)e1 \\
		3 & -1.68233(19)e4 & -7.7922(9)e3 & -2.74402(29)e3 & -5.340(6)e2 & -2.91260(20)e2 \\
		4 & -6.8157(12)e5 & -2.4464(5)e5 & -6.0962(10)e4 & -6.867(20)e3 & -3.0556(4)e3 \\
		5 & -3.0924(7)e7 & -8.6051(21)e6 & -1.5193(5)e6 & -9.94(4)e4 & -3.6053(6)e4 \\
		6 & -1.5029(8)e9 & -3.2427(16)e8 & -4.0588(20)e7 & -1.547(5)e6 & -4.5835(9)e5 \\
		7 & -7.650(6)e10 & -1.2799(9)e10 & -1.1360(7)e9 & -2.525(9)e7 & -6.1169(20)e6 \\
		8 & -4.027(8)e12 & -5.223(6)e11 & -3.2872(27)e10 & -4.269(20)e8 & -8.455(4)e7 \\
		9 & -2.174(12)e14 & -2.185(6)e13 & -9.753(18)e11 & -7.40(8)e9 & -1.1998(10)e9 \\[1em]

		 & \multicolumn{1}{c}{3.5} & \multicolumn{1}{c}{4} & \multicolumn{1}{c}{5} & \multicolumn{1}{c}{6} & \multicolumn{1}{c}{8} \\
		\midrule
		2 & -3.22939(21)e1 & -2.36263(14)e1 & -1.6830(4)e1 & -1.43161(8)e1 & -1.26312(12)e1 \\
		3 & -2.30987(23)e2 & -1.37616(14)e2 & -7.6763(21)e1 & -5.7613(8)e1 & -4.6010(11)e1 \\
		4 & -2.2420(4)e3 & -1.12591(22)e3 & -5.2550(29)e2 & -3.6656(9)e2 & -2.7981(14)e2 \\
		5 & -2.4454(6)e4 & -1.03022(25)e4 & -3.965(4)e3 & -2.5227(10)e3 & -1.7989(15)e3 \\
		6 & -2.8767(11)e5 & -1.0206(4)e5 & -3.264(6)e4 & -1.9154(16)e4 & -1.2913(22)e4 \\
		7 & -3.5538(17)e6 & -1.0617(6)e6 & -2.815(6)e5 & -1.5170(16)e5 & -9.592(26)e4 \\
		8 & -4.5484(28)e7 & -1.1458(10)e7 & -2.529(12)e6 & -1.2546(28)e6 & -7.49(4)e5 \\
		9 & -5.978(7)e8 & -1.2697(21)e8 & -2.336(19)e7 & -1.066(11)e7 & -6.0(4)e6 \\[1em]

		& \multicolumn{1}{c}{10} \\
		\midrule
		2 & -1.21853(27)e1 \\
		3 & -4.318(5)e1 \\
		4 & -2.6012(16)e2 \\
		5 & -1.646(8)e3 \\
		6 & -1.164(5)e4 \\
		7 & -8.46(10)e4 \\
		8 & -6.3(13)e5 \\
		9 & -4.0(25)e6 \\

		\bottomrule
		\end{tabular}

	% \end{ruledtabular}
	\egroup
	\end{table}
	\endgroup

\begingroup
\squeezetable
\begin{table}[tb]
	\caption{Coefficients $c_k$ of the gap $\Delta$ for each order $k$ in the antiferromagnetic lrTFIM on the square lattice (\(\vec k = \left(\pi,\pi\right)^T\)).}
	\label{tab:coeff_square_af}
	\centering
	\bgroup
	\setlength\tabcolsep{.8em}
	% \begin{ruledtabular}
		% \newcolumntype{e}[1]{D{e}{\mathrm{e}}{#1} }
		% \end{tabular}
		\begin{tabular}{cSSSSS}
		\toprule
		\diagbox{$k$}{$\alpha$} & \multicolumn{1}{c}{2.5} & \multicolumn{1}{c}{3} & \multicolumn{1}{c}{3.5} & \multicolumn{1}{c}{4} & \multicolumn{1}{c}{5} \\
		\midrule
		2 & 2.1515(10)e0 & 1.1586(6)e0 & 3.851(5)e-1 & -2.589(4)e-1 & -1.2822(4)e0 \\
		3 & 2.3853(16)e1 & 1.5072(5)e1 & 1.13546(22)e1 & 9.6405(15)e0 & 8.6559(13)e0 \\
		4 & 2.495(5)e2 & 1.0271(6)e2 & 5.1616(22)e1 & 2.7367(13)e1 & 4.035(9)e0 \\
		5 & 3.433(10)e3 & 1.0238(7)e3 & 4.5085(26)e2 & 2.5320(11)e2 & 1.3571(7)e2 \\
		6 & 5.338(27)e4 & 1.0776(13)e4 & 3.6259(27)e3 & 1.5870(13)e3 & 4.094(6)e2 \\
		7 & 9.14(10)e5 & 1.2364(20)e5 & 3.292(4)e4 & 1.2761(14)e4 & 3.979(6)e3 \\
		8 & 1.668(24)e7 & 1.492(4)e6 & 3.072(6)e5 & 9.846(18)e4 & 2.028(7)e4 \\
		9 & 3.24(13)e8 & 1.878(9)e7 & 2.993(11)e6 & 8.136(30)e5 & 1.552(10)e5 \\[1em]

		 & \multicolumn{1}{c}{6} & \multicolumn{1}{c}{8} & \multicolumn{1}{c}{10} \\
		\midrule
		2 & -2.0407(4)e0 & -3.0015(6)e0 & -3.4982(6)e0 \\
		3 & 8.8971(16)e0 & 1.00499(27)e1 & 1.09394(27)e1 \\
		4 & -8.110(12)e0 & -2.1665(20)e1 & -2.8654(24)e1 \\
		5 & 1.1595(6)e2 & 1.3041(14)e2 & 1.4982(20)e2 \\
		6 & 3.64(6)e1 & -2.974(11)e2 & -4.676(21)e2 \\
		7 & 2.509(5)e3 & 2.513(8)e3 & 2.973(17)e3 \\
		8 & 5.22(6)e3 & -4.57(10)e3 & -9.27(19)e3 \\
		9 & 6.97(7)e4 & 5.66(9)e4 & 7.0(13)e4 \\

		\bottomrule
		\end{tabular}

	% \end{ruledtabular}
	\egroup
	\end{table}
	\endgroup

\begingroup
\squeezetable
\begin{table}[tb]
	\caption{Coefficients $c_k$ of $\Delta$ for each order $k$ in the antiferromagnetic lrTFIM on the triangular lattice (\(\vec k =  \left( \frac23\pi,-\frac23\pi\right)^T\)).}
	\label{tab:coeff_triangular_af}
	\centering
	\bgroup
	\setlength\tabcolsep{.8em}
	% \begin{ruledtabular}
		% \newcolumntype{e}[1]{D{e}{\mathrm{e}}{#1} }
		% \end{tabular}
		\begin{tabular}{cSSSSSS}
		\toprule
		\diagbox{$k$}{$\alpha$} & \multicolumn{1}{c}{2.25} & \multicolumn{1}{c}{2.5} & \multicolumn{1}{c}{3} & \multicolumn{1}{c}{3.5} & \multicolumn{1}{c}{4} & \multicolumn{1}{c}{5} \\
		\midrule
		2 & 4.942(5)e0 & 4.3943(16)e0 & 3.6574(9)e0 & 3.1636(4)e0 & 2.8010(4)e0 & 2.3043(5)e0 \\
		3 & 5.286(13)e1 & 3.9088(27)e1 & 2.5376(10)e1 & 1.8759(4)e1 & 1.49914(26)e1 & 1.11057(29)e1 \\
		4 & 8.76(6)e2 & 5.148(7)e2 & 2.4772(15)e2 & 1.5071(5)e2 & 1.04611(30)e2 & 6.3726(25)e1 \\
		5 & 1.85(4)e4 & 8.358(18)e3 & 2.9021(26)e3 & 1.4511(6)e3 & 8.870(4)e2 & 4.7002(29)e2 \\
		6 & 4.55(22)e5 & 1.547(6)e5 & 3.778(6)e4 & 1.5324(10)e4 & 8.172(5)e3 & 3.701(4)e3 \\
		7 & 1.26(22)e7 & 3.143(19)e6 & 5.295(11)e5 & 1.7256(17)e5 & 7.990(7)e4 & 3.092(5)e4 \\
		8 & 3.7(20)e8 & 6.83(6)e7 & 7.830(21)e6 & 2.036(4)e6 & 8.154(13)e5 & 2.680(9)e5 \\
		9 & 1(7)e10 & 1.574(27)e9 & 1.207(7)e8 & 2.492(7)e7 & 8.606(28)e6 & 2.393(17)e6 \\[1em]

		 & \multicolumn{1}{c}{6} & \multicolumn{1}{c}{8} & \multicolumn{1}{c}{10} \\
		\midrule
		2 & 1.9948(4)e0 & 1.6816(5)e0 & 1.5645(6)e0 \\
		3 & 9.3368(21)e0 & 8.0165(27)e0 & 7.654(4)e0 \\
		4 & 4.7125(16)e1 & 3.5224(20)e1 & 3.1888(23)e1 \\
		5 & 3.2851(20)e2 & 2.4022(22)e2 & 2.1819(27)e2 \\
		6 & 2.3943(22)e3 & 1.646(4)e3 & 1.467(5)e3 \\
		7 & 1.862(4)e4 & 1.219(4)e4 & 1.078(6)e4 \\
		8 & 1.496(5)e5 & 9.27(7)e4 & 8.03(20)e4 \\
		9 & 1.239(19)e6 & 7.2(4)e5 & 6.3(12)e5 \\

		\bottomrule
		\end{tabular}

	% \end{ruledtabular}
	\egroup
	\end{table}
	\endgroup

%====================================================================================================================
% Extrapolation
%====================================================================================================================
\section{Extrapolation}
Once the energy gap is given as a power series (see last section), we perform standard DlogPad\'{e} extrapolations. We refer to the literature for a general review of this topic, as for example given in Ref.~\onlinecite{Guttmann1989}. Here we give specific information which is relevant for the particular extrapolation we performed in the main body of the manuscript. 

Our series are all of the form
\begin{align}
F(\lambda)=\sum_{k\geq 0}^{k_{\mathrm{max}}} c_k \lambda^k=c_0+c_1\lambda+c_2\lambda^2+\dots c_{k_{\mathrm{max}}}\lambda^{k_{\mathrm{max}}},
\end{align}
with $\lambda\in \mathbb{R}$ and $c_k \in \mathbb{R}$. If one has power-law behavior near a critical value $\lambda_{\rm c}$, the true physical function $\tilde{F}(\lambda)$ close to $\lambda_{\rm c}$ is given by
\begin{align}
\tilde{F}(\lambda)\approx \left(1-\frac{\lambda}{\lambda_{\rm c}}\right)^{-\theta} A(\lambda),
\end{align}
where $\theta$ is the associated critical exponent. If $A(\lambda)$ is analytic at $\lambda=\lambda_{\rm c}$, we can write
\begin{align}
\label{eq:Ftilde}
\tilde{F}(\lambda)\approx \left(1-\frac{\lambda}{\lambda_{\rm c}}\right)^{-\theta}A\Big|_{\lambda=\lambda_{\rm c}}\left(1+\mathcal{O}\left(1-\frac{\lambda}{\lambda_{\rm c}}\right)\right).
\end{align}
Near the critical value $\lambda_{\rm c}$, the logarithmic derivative is then given by
\begin{align}
\tilde{D}(\lambda)&:=\frac{\text{d}}{\text{d}\lambda}\ln{\tilde{F}(\lambda)}\label{dx}\\
&\approx \frac{\theta}{\lambda_{\rm c}-\lambda}\left\{ 1+ \mathcal{O}(\lambda-\lambda_{\rm c})\right\}\nonumber.
\end{align}
In the case of power-law behavior, the logarithmic derivative $\tilde{D}(\lambda)$ is therefore expected to exhibit a single pole at $\lambda\equiv\lambda_{\rm c}$.

The latter is the reason why so-called DlogPad\'{e} extrapolation is often used to extract critical points and critical exponents from high-order series expansions. DlogPad\'e extrapolants of $F(\lambda)$ are defined by
\begin{align}
\label{eq:dlogP1}
dP[L/M]_F(\lambda)=\exp\left(\int_{0}^\lambda P[L/M]_{D}\,\,\text{d}\lambda'\right)
\end{align}
and represent physically grounded extrapolants in the case of a second-order phase transition. Here $P[L/M]_{D}$ denotes a standard Pad\'e extrapolation of the logarithmic derivative
\begin{align}
\label{eq:dlogP2}
P[L/M]_{D}:=\frac{P_L(\lambda)}{Q_M(\lambda)}=\frac{p_0+p_1\lambda+\dots + p_L \lambda^L}{q_0+q_1\lambda+\dots q_M \lambda^M}\quad,
\end{align}
with $p_i\in \mathbb{R}$, $q_i \in \mathbb{R}$, and $q_0=1$. Additionally, $L$ and $M$ have to be chosen so that $L+M\leq k_{\mathrm{max}}-1$. Physical poles of $P[L/M]_{D}(\lambda)$ then indicate critical values $\lambda_{\rm c}$ while the corresponding critical exponent of the pole $\lambda_{\rm c}$ can be deduced by
\begin{align}
\theta\equiv\left.\frac{P_L(\lambda)}{\frac{\text{d}}{\text{d}\lambda} Q_M(\lambda)}\right|_{\lambda=\lambda_{\rm c}} \label{extract_exponent}.
\end{align}
If the exact value (or a quantitative estimate from other approaches) of $\lambda_{\rm c}$ is known, one can obtain better estimates of the critical exponent by defining
\begin{align*}
\theta^*(\lambda)&\equiv(\lambda_{\rm c}-\lambda)D(\lambda)\\
&\approx \theta+\mathcal{O}(\lambda-\lambda_{\rm c}),
\end{align*}
where $D(\lambda)$ is given by Eq.~\eqref{dx}. Then
\begin{align}
P[L/M]_{\theta^*}\big|_{\lambda=\lambda_{\rm c}}=\theta \label{biasnue}
\end{align}
yields a (biased) estimate of the critical exponent.

In the ferromagnetic case at the upper critical $\alpha=10/3$ for two dimensions, the lrTFIM displays multiplicative corrections close to the quantum critical point so that one expects the following critical behavior
\begin{align}
\bar{F}(\lambda)\approx \left(1-\frac{\lambda}{\lambda_{\rm c}}\right)^{-\theta} \left(\ln\left( 1-\frac{\lambda}{\lambda_{\rm c}}\right)\right)^{p} \bar{A}(\lambda),
\end{align}
where $\lambda_{\rm c}$ ($\theta$) is the associated critical point (exponent) as before while $p$ yields the exponent of multiplicative logarithmic corrections. Clearly, the extraction of $p$ from a high-order series expansion is very demanding. The only reasonable approach is to bias the extrapolation by fixing $\theta$. In our case the critical exponent $\theta$ at \(\alpha=10/3\) is given by the well-known mean-field value $1/2$.

Assuming again that the function $\bar{A}(\lambda)$ is analytic close to $\lambda_{\rm c}$, Eq.~\eqref{eq:Ftilde} transforms into
\begin{eqnarray}
\label{eq:Fbar}
\bar{F}(\lambda)&\approx& \left(1-\frac{\lambda}{\lambda_{\rm c}}\right)^{-\theta} \left(\ln\left( 1-\frac{\lambda}{\lambda_{\rm c}}\right)\right)^{p}\bar{A}|_{\lambda=\lambda_{\rm c}}\nonumber \left(1+\mathcal{O}\left(1-\frac{\lambda}{\lambda_{\rm c}}\right)\right).
\end{eqnarray}
and the logarithmic derivative Eq.~\eqref{dx} becomes
\begin{align}
\bar{D}(\lambda)&\approx \frac{\theta}{\lambda_{\rm c}-\lambda} + \frac{-p}{\ln\left(1-\lambda/\lambda_{\rm c}\right)\left(\lambda_{\rm c}-\lambda\right)} + \mathcal{O}\left(\lambda-\lambda_{\rm c}\right)\nonumber.
\end{align}
One can then estimate the multiplicative logarithmic correction $p$ by defining
\begin{align*}
  p^{*}(\lambda)&\equiv -\ln\left( 1-\lambda/\lambda_{\rm c}\right) \left[  \left( \lambda_{\rm c}-\lambda\right) D(\lambda)-\theta \right]\\
           &\approx p +\mathcal{O}(\lambda-\lambda_{\rm c}),
\end{align*}
and by performing Pad\'{e} extrapolants of this function
\begin{align}
P[L/M]_{p*}\big|_{\lambda=\lambda_{\rm c}}=p \label{biasp}\quad .
\end{align}

For our results we study the possible combinations of the order of the numerator and denominator polynomial $L$ and $M$. We sort them into the families $[M,M-2]$, $[M,M+2]$, $[M,M-1]$, $[M,M+1]$, and $[M,M]$ and analyze their convergence. We then take the highest order Pad\'e extrapolants of converging families and calculate the mean value and the standard deviation of the critical $\lambda_{\mathrm c}$ and the exponents $z\nu$. These values are shown in the main paper's results.

\end{document}